\documentstyle[aps,prl,graphicx,floats]{revtex}
\begin{document}
\draft
\twocolumn[\hsize\textwidth\columnwidth\hsize\csname @twocolumnfalse\endcsname

\title{ Quantum tunneling between paramagnetic and superconducting
states of a nanometer-scale superconducting grain placed in a
magnetic field.}

\author{A.V. Lopatin  and  V.M. Vinokur}
\address{
Materials Science Division, Argonne National Laboratory,
Argonne, Illinois 60439
}
\date{\today}
\maketitle

\begin{abstract}
 We consider the process of quantum tunneling between the superconducting
and paramagnetic states of
a nanometer-scale superconducting grain placed in a magnetic
field.
The grain is supposed to be coupled via tunneling junction
to a normal metallic contact that plays a role of the spin
reservoir. Using the instanton method we find the probability of
the quantum tunneling process and express it in terms of the
applied magnetic field, order parameter of the superconducting
grain and conductance of the tunneling junction between the grain
and  metallic contact.

\end{abstract}

\vskip2pc]

\bigskip

Recent advances in manufacturing of small electronic devices posed
several questions in the theory of nanometer-size superconductors
related to their prospective applications \cite{Grains_Experiment}.
One of the key issues
is the behavior of ultra-small superconductors, of the dimensions
much less than the coherence length $\xi$ in the
presence of a magnetic field. Of a special interest from both
fundamental and practical point of view are the dynamical
processes of switching between the superconducting and
paramagnetic states.   Small grains can experience spontaneous
transitions due to quantum fluctuations. In this Letter we
investigate the process of quantum tunneling between the
superconducting and paramagnetic states of an ultra-small
superconductor.

In geometrically restricted superconductors the Zeeman effect
suppresses superconductivity at magnetic fields exceeding
$H_{spin}= \Delta/ \sqrt{2}\, \mu_B$ where $\Delta$ is the order
parameter \cite{Films_theory,Films_1,Films_2}.  The orbital effect
in small superconductors, which usually plays a major role, was
discussed in a pioneering work by Larkin \cite{larkin}: In a dirty
spherical grain, $\Delta\tau\ll 1$, $\tau$ is the normal
excitation relaxation time, the orbital effect leads to
suppression of superconductivity at
$H_{orb}\sim\Phi_0/(r\sqrt{D/\Delta})$, where $\Phi _0=hc/2e$, $D$
is the diffusion coefficient, and $r$ is the radius of the grain.

Tunneling switching effects appear in the paramagnetic limit
$H_{spin}/H_{orb}\sim \sqrt{ \Delta /d } \, \sqrt{l/r}\ll 1,$
where $l=v_F\tau$ is the mean free path, and $d $ is the mean
energy level spacing.
   In the ballistic case ($l\simeq r$)
this limit is achieved only if  $\Delta <d$ corresponding to the
strongly fluctuating regime\cite{Anderson}. However in a platelet
geometry with  magnetic field applied along the film ${{H_{spin}
}/{H_{orb}  }} \sim \sqrt{\Delta/ d} \, \sqrt{lb/S}$, where $b$ is
the thickness and $S$ is the area of the sample; thus the
paramagnetic limit can be easily achieved along with condition
$\Delta\gg d$ even in the ballistic case if the ratio $b/\sqrt{S}$
is sufficiently small.

We focus on the case where the Zeeman effect dominates the orbital
one while the order parameter $\Delta$ is larger than the mean
level spacing $d$  and investigate the process of quantum
tunneling between superconducting and paramagnetic states. Since
these two states are characterized by different values of the
total spin, such quantum tunneling process can occur only in
presence of processes allowing for non-conservation of the total
spin.  We consider the grain weakly coupled with a normal metallic
lead (plate) that plays a role of a spin bath. Our final result
for the probability of the quantum tunneling between the
superconducting and paramagnetic states is
     \begin{equation}
           \label{answer}P\sim {\rm exp}\;\left[ N\;\ln (\,\beta \delta EG/\Delta
            _0\,)\,\right]
          \end{equation}
where the numerical coefficient $\beta \approx 1.1,$ \ the factor
$N$ is the number of polarized electrons in the paramagnetic state
of the grain, $G$ is the conductance of the tunneling junction
between the grain and the metallic lead measured in units $e^2/h$,
$\Delta_0 $ is the order parameter of the grain in the
superconducting state and\ $N\delta E$ is the total energy
difference between the grain's superconducting and paramagnetic
states. The factor $N$ is related to the applied magnetic filed
$H$ and the
average density of states $\nu $ as \ $N=2 \nu \mu _B H $ with \ $\mu _B=|e|%
\bar h/2mc$.\ The energy difference per one state $\delta E$ is
related to the magnetic field deviation $\delta H$ from the
magnitude at which the the thermodynamic phase transition occurs
via $\delta E=\mu _B\delta H$. The result (\ref{answer}) is valid
as long as $\delta E/\Delta \ll 1$, the general case $\Delta \sim
\delta E$ is more complicated and we leave it for future study.
The meaning of our answer (\ref{answer}) can be understood as
follows. The probability of a single tunneling process is $
(\delta E/\Delta_0)G $, then the probability of having $N$ of them
simultaneously in order to transit to paramagnetic state is simply
$ [(\delta E/\Delta_0)G]^N $.

 {\it The model.} We consider a system of
a superconducting grain and a metallic plate coupled via weak
tunneling
\begin{equation}
\label{totalham}\hat H=\hat H_g+\hat H_M+\sum_{k,k^{\prime
}}\,t_{k k^{\prime }}[\hat \psi _{k\sigma }^{\dagger }\hat d_{k^{\prime
}\sigma }+\hat d_{k^{\prime }\sigma }^{\dagger }\hat \psi _{k\sigma })].
\end{equation}
where $\hat H_g$ and $\hat H_M$ are the Hamiltonians of the grain and
metallic plate, $\hat \psi ^{\dagger }(\hat \psi )$ and $\hat d^{\dagger }(%
\widehat{d})$ are the creation (annihilation) operators of
electrons of the grain and of the metal respectively and
$t_{kk^{\prime }}$ is the electron tunneling matrix element
between the grain and metal. Electrons of metallic plate are
described by the free-fermion Hamiltonian
$$
\hat H_M=\sum_{k^{\prime }}\hat d_{k^{\prime }\sigma }^{\dagger }\,\zeta
_{\sigma k^{\prime }}\,\hat d_{k^{\prime }\sigma }.
$$
while the grain is described by the BCS model
\begin{eqnarray}
H_{g}=\sum_{k}\psi _{k}^{\dagger }&[&\xi_k -h\sigma _{z}\,]\psi
_{k} \nonumber \\
&-&\lambda \sum_{k1,k2}\psi _{k1\uparrow }^{\dagger
}\psi _{k1\downarrow }^{\dagger }\psi _{k2\downarrow }\psi
_{k2\uparrow ,}  \label{Ham1}
\end{eqnarray}
where $\xi _{k}$ are the exact eigenvalues (measured with respect
to the chemical potential) of the noninteracting Hamiltonian,
$\lambda$ is the coupling constant, and the magnetic field $h$,
pointing along the $z-$ axis, is measured in the energy units
$h=\mu _{B}H.$

{ \it Quantum tunneling.} The amplitude of the tunneling process
between the initial and final states is
\begin{equation}
\label{ampl}A=<f\,\,|\;T_{t}\;e^{-i\int_{t_{i}}^{t_{f}}\hat{H}(t)\,d t
}\;|\,\,i>,
\end{equation}
where the Hamiltonian $\hat{H}(t)$ is written in the interaction
representation: the noninteracting part includes the metal Hamiltonian $%
H_{M} $ and the first (free-fermion) term of grain Hamiltonian
(\ref{Ham1}) while the interaction part includes the BCS
interaction and electron tunneling between the grain and metal. We
will consider a process of quantum tunneling from pure
superconducting state (initial state) to the paramagnetic state
with order parameter $\Delta =0$. The electron tunneling between
the grain and metal will be treated as a perturbation assuming
that the tunneling matrix elements $t_{k k\prime }$ are small. The
paramagnetic state has a nonzero total spin $S$ which is formed by
the polarized electrons with $|\xi_k|<\tilde \xi$ such that
$S=\nu\tilde\xi.$ During the quantum tunneling process the spin of
the grain increases from zero to $S,$ thus there must be $2S$
electron tunneling processes between the grain and metal and first
nonzero contribution in expansion  of Eq.(\ref {ampl}) in
tunneling element emerges only in $N=2S$ order. It is clear that
the
paired states which are destroyed by the electron tunneling are those with $|\xi |<%
\tilde{\xi}.$ Expanding Eq.(\ref{ampl}) it $t$ we have
\begin{eqnarray}
A =<f\,|\;\prod_{|\xi _{k}|<\tilde{\xi}}\int
dt_{k}\,T_{t}\;e^{-i\int_{t_{i}}^{t_{f}}\hat{H}(t\,)dt}\sum_{k^{\prime
}}t_{kk^{\prime }}  \nonumber  \label{ampl1} \\ \times \lbrack
\hat{\psi}_{k\,\sigma }^{\dagger }(t_{k})\hat{d}_{k^{\prime
}\sigma }(t_{k})+\hat{d}_{k^{\prime }\sigma }^{\dagger }(t_{k})\hat{\psi}%
_{k\,\sigma }(t_{k})]\;|\,i >. \label{ampl1}
\end{eqnarray}

In the absence of coupling between the grain and metal the
initial and final states of the system are the products of the
corresponding initial and final  states of the grain and metal
$|i>=|i_{G}>|i_{M}>,\;|f>=|f_{G}>|f_{M}>,$ thus in the leading order in
tunneling matrix element the
quantum mechanical average of operators $\hat{d}$ in (\ref{ampl1})
can be directly implemented. Since the total spin of the metallic
plate decreases during the quantum tunneling process the only
relevant matrix elements are those that correspond to creation of
electrons with spin down and annihilation of electrons with spin up
\begin{eqnarray}
&<&f_{M}|\,d_{k\downarrow }^{\dagger }(t)\,|i_{M}>=e^{i\zeta _{k\downarrow
}\,t},\;\;\;\;\;\;\;   \zeta _{k\downarrow }>0  \nonumber  \\
&<&f_{M}|\,d_{k\uparrow }(t)\,|i_{M}>=e^{i|\zeta _{k\uparrow
}|\,t},\;\;\;\;\;\zeta _{k\uparrow }<0. \nonumber
\end{eqnarray}
The initial state of the metal is the Fermi sea while its final
state having $N$ electron-hole excitations is characterized by the
set of excitation energies $\zeta _{k^{\prime }\alpha }^{\{p\}}$
where the index  $k^{\prime }$ labels the quantum number of the
excitation $p.$ Now assuming that tunneling matrix elements
are index independent $t_{k,k^\prime} = t$ the amplitude can be written as
\begin{eqnarray}
A &=&t^{N}<f_{G}|T_{t}\prod_{|\xi _{k}|<\tilde{\xi}}\int
dt_{k}e^{-i\int_{t_{i}}^{t_{f}}\hat{H}(t)dt}\sum_{{\rm
Per}\;p(k)}e^{i\zeta _{k^{\prime }}^{\{p(k)\}}t_{k}}  \nonumber
\\ \label{ampl2} &&\times \lbrack \hat{\psi}_{k\uparrow }^{\dagger
}(t_{k})\,\theta (-\zeta _{k^{\prime
}}^{\{p(k)\}})+\hat{\psi}_{k\downarrow }(t_{k})\,\theta (\zeta
_{k^{\prime }}^{\{p(k)\}})]\;\,|i_{G}>
\end{eqnarray}
where the sum goes over all excitation permutations $p(k).$ Since
we consider only the case $\delta E<<\Delta$ we can simplify the
problem assuming that the energy of a typical excitation in the
metal is much less than $\Delta.$ In this case we can neglect
the energies $\zeta _{k\prime }^{\{p\}}$ in (\ref{ampl2}) and
using the Heisenberg representation we write the amplitude
(\ref{ampl2}) as
\begin{eqnarray}
A =N!\,t^{N}<f_{G}|\,\prod_{|\xi _{k}|<\tilde{\xi}}\int dt_{k}\,[\hat{\psi}%
_{k\uparrow }^{\dagger }(t_{k})\,\theta (-\zeta _{k^{\prime }}^{\{p(k)\}})
\nonumber \\
+\hat{\psi}_{k\downarrow }(t_{k})\,\theta (\zeta _{k^{\prime
}}^{\{p(k)\}})]\,|i_{G} >,  \label{amplA}
\end{eqnarray}
where a specific permutation $p(k)$ was chosen.

The probability $P$ of the tunneling process is obtained by
integrating of $A^* A $ over all possible quantum states of the
metal
\begin{equation}
\label{total_prob}P\sim {\frac{\nu_M^N}{{N!}}}\,  \int
d\zeta^{\{1\}}..d \zeta^{\{N\}} \, A^* A\, \delta(N\, \delta E-\sum_p
|\zeta^{\{p\}}|),
\end{equation}
where $\nu_M$ is the density of states of the metal.

\begin{figure}[ht]
\hspace{0.3cm}
\includegraphics[width=2.8in]{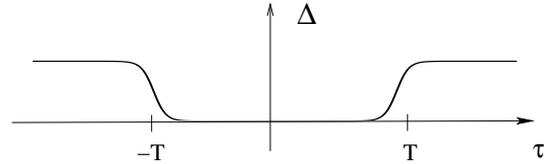}
\caption{ Dependence of the order parameter $\Delta$ on time $\protect\tau.$
The region $\protect\tau>0$ is the mirror reflection of the ``physical''
region $\protect\tau<0.$ The superconducting state corresponds to the regions
$|\protect\tau|> T$ while the paramagnetic one to $|\protect\tau|< T.$ }
\end{figure}

To find the value of $A$ we shall use the instanton method turning to use
the Euclidean time $t\rightarrow -i\tau .$ Taking initial and final times as
$\tau _{i}=-\infty ,\tau _{f}=0$ we present the amplitude $A$ as
$$
A=N!\,t^{N}\,<f_{G}|\,S(0,-\infty )\,|i_{G}>
$$
with the evolution operator
\begin{eqnarray}
S(\tau _{2},\tau _{1}) &=&\prod_{|\xi _{k}|<\tilde{\xi}}\int_{\tau
_{1}}^{\tau _{2}}d\tau _{k}\,[\hat{\psi}_{k\uparrow }^{\dagger }(\tau
_{k})\,\theta (-\zeta _{k^{\prime }}^{\{p(k)\}})  \nonumber  \label{ev_op_S}
\\
&&+\hat{\psi}_{k\downarrow }(\tau _{k})\,\theta (\zeta _{k^{\prime
}}^{\{p(k)\}})].
\end{eqnarray}
In principle, the instanton method should be applied to the
amplitude $A$ directly, but it is more convenient to consider the
product $A^{\ast }A$ presenting $A^{\ast }$ as $A^{\ast
}=N!\,t^{N}<i_{G}|S^{\dagger }(\infty ,0)|f_{G}>$ and writing
the product $A^{\ast }A$ as
\begin{eqnarray}
A^{\ast }A &=&<i_{G}|\,S^{\dagger }(\infty
,0)\,|f_{G}>\,<f_{G}|\,S(0,-\infty )|i_{G}>  \nonumber \\
&=&<i_{G}|S^{\dagger }(\infty ,0)\,S(0,-\infty )\,|i_{G}>, \nonumber
\end{eqnarray}
where it was used that by construction the Euclidean evolution
operator $S$
brings the grain from initial $|i_{G}>$ to the final $|f_{G}>$ state: $%
S\,|i_{G}>=|f_{G}>.$ Now the instanton process has the following
structure (see Fig.1.): the evolution begins at $\tau =-\infty $
from the superconducting state, then at $\tau \approx -T$ the
system turns into the paramagnetic state and stays there till it
turns back to the superconducting state at $\tau \approx T.$ The
artificial part of the process ($\tau >0$) is the mirror reflection of
the ``physical'' process with $\tau <0.$ The advantage of this
representation is that now the new initial ($\tau =-\infty $) and
final ($\tau =\infty $) states are identical and therefore we can
use the convenient functional representation for $A^{\ast }A$
\begin{eqnarray}
A^{\ast }\,A &=&[N!\,t^{N}]^{2}\int D\psi ^{\dagger }
\,D\psi \,e^{\int \,{\cal L}dt}\prod_{k}\int d\tau _{1k}d\tau _{2k}
\nonumber \\
&&\times \Bigr[\psi _{k\uparrow }(\tau _{1k})\,\psi _{k\uparrow }^{\dagger
}(\tau _{2k})\,\theta (-\zeta _{k^{\prime }}^{\{p(k)\}})  \nonumber \\
&&\;\;\;\;\;\;\;\;\;\;\;\;\;\;\;+\psi _{k\downarrow }^{\dagger }(\tau
_{1k})\,\psi _{k\downarrow }(\tau _{2k})\theta (\zeta _{k^{\prime
}}^{\{p(k)\}})\Bigl],  \label{A*A}
\end{eqnarray}
with the Lagrangian
\begin{eqnarray}
{\cal L} =-\frac{1}{\lambda}\Delta ^{\ast }\Delta -\sum_{k}\psi
_{k}^{\dagger }[\partial _{\tau }+\xi _{k}-h\,\sigma _{z}\,]\psi
_{k}  \nonumber
\\
 \label{lagrp} -\Delta \,\psi _{k\uparrow
}^{\dagger }\psi _{k\downarrow }^{\dagger }-\Delta ^{\ast }\,\psi
_{k\downarrow }\psi _{k\uparrow }.
\end{eqnarray}
The integration over the fermionic fields in (\ref{A*A}) should be
implemented exactly while the assumed integration over $\Delta
(\tau )$ will be done with the saddle point accuracy. Now taking
integrals in (\ref {total_prob}) for the probability we obtain
\begin{eqnarray}
P &=&e^{N\ln (N\,\delta E\,t^{2}\,\nu _{M})}\int D\psi ^{\dagger
}\,D\psi \,\prod_{k}\int d\tau _{1k}d\tau _{2k}  \nonumber
\\
&&\times e^{\int \,{\cal L}d\tau }[\psi _{k\uparrow }(\tau _{1k})\,\psi
_{k\uparrow }^{\dagger }(\tau _{2k})+\psi _{k\downarrow }^{\dagger }(\tau
_{1k})\,\psi _{k\downarrow }(\tau _{2k})], \nonumber
\end{eqnarray}
and finally integrating over the fermionic fields we obtain
\begin{eqnarray}
\ln P =N\,\ln (N\,\delta E\,t^{2}\,\nu _{M})+\sum_{k}{\rm
Tr}\,\ln [\partial _{\tau }+{\cal H}_{k}]  \nonumber \\
\label{final_prob}-{\frac{1}{g}}\int d\tau \,\Delta ^{\ast }(\tau
)\Delta (\tau )+\sum_{k}\ln Z_k, \label{probabillity}
\end{eqnarray}
where  $Z_k={\rm Tr}\int d\tau _{1}d\tau
_{2}\;\hat{G}_{1k}(\tau _{1},\tau _{2}),$
\begin{equation}
{\cal H}_{k}(\tau )=\left[
\begin{array}{cc}
\xi _{k}-h & \Delta (\tau ) \\
\Delta ^{\ast }(\tau ) & -\xi _{k}-h
\end{array}
\right] ,    \label{matrix_H}
\end{equation}
and the matrix Green function
\begin{equation}
\hat{G}_{1k}(\tau _{1},\tau _{2})=\left[
\begin{array}{cc}
G_{1k}(\tau _{1},\tau _{2}) & F_{1k}(\tau _{1},\tau _{2}) \\
F_{1k}^{\dagger }(\tau _{1},\tau _{2}) & \bar{G}_{1k}(\tau
_{1},\tau _{2})
\end{array}
\right] .
\end{equation}
is defined by the equation
\begin{equation}
\label{green_func_eq}\lbrack \partial _{\tau 1}+{\cal H}_{k}(\tau _{1})]\;%
\hat{G}_{1k}(\tau _{1},\tau _{2})=\delta (\tau _{1}-\tau _{2}).
\end{equation}

{\it Instanton equations.} To find the instanton equations we take the
functional derivative of Eq.(\ref{final_prob}) with respect to $\Delta^*$
obtaining
\begin{equation}
\label{self_cons_delta}{\frac{1}{\lambda }} \Delta(\tau)= \sum_k
f_{1k}(\tau)+f_{2k}(\tau),
\end{equation}
where the Green function $f_1(\tau)=F_1(\tau,\tau)$ emerges form the
variation derivative of ${\rm Tr} \ln[\partial_\tau +{\cal H}_k] $ in (\ref
{final_prob}) and $f_2$ emerges from the functional derivative of the Green
function $\hat G_1 $
\begin{equation}
\label{f2_var}f_2(\tau)=Z_k^{-1} {\frac{{\delta }}{{\delta \Delta^*(\tau) }}}
\int\, d \tau_1 d \tau_2\, {\rm Tr} \, G_{1k}(\tau_1,\tau_2).
\end{equation}

   Combining Eq.(\ref{green_func_eq}) with the same equation written in the transposed form
one can find the equation that defines the function $f_1$ in terms of only diagonal Green
functions
\begin{equation}
\label{eilenberger}\partial _\tau \hat g_{1k}(\tau )+[{\cal H}_{0k}(\tau ),\hat g_{1k}(\tau )]=0,
\end{equation}
where
\begin{equation}
\label{definition_g1}\hat g_{1k}(\tau )=\left[
\begin{array}{cc}
g_{1k}(\tau ) & f_{1k}(\tau ) \\
f_{1k}^{\dagger }(\tau ) & \bar g_{1k}(\tau )
\end{array}
\right] =\hat G_{1k}(\tau ,\tau ),
\end{equation}
and ${\cal H}_0$ is given by Eq.(\ref{matrix_H}) with $h=0.$
Writing Eq.(\ref{eilenberger}) in components we get
\begin{eqnarray}\nonumber
\partial _{\tau }\tilde{g}_{1k}+\Delta f_{1k}^{\dagger }-\Delta ^{\ast
}f_{1k} &=&0,   \\
\nonumber\partial _{\tau }f_{1k}+2\xi_k f_{1k}-2\Delta \tilde{g}_{1k} &=&0,  \\
-\partial _{\tau }f_{1k}^{\dagger }+2\xi_k f_{1k}^{\dagger }-2\Delta ^{\ast }%
\tilde{g}_{1k} &=&0, \label{eqfd} \\
\label{eqgz}\partial _\tau s_{z1k}=0,
\end{eqnarray}
where the variables $\tilde g_1=[g_{1k}-\bar g_{1k}]/2,$ $\ s_{z1k}=-[g_{1k}+
\bar g_{1k}]/2$ were introduced instead of components $g_1$ and
$\bar g_1.$ Eqs.(\ref{eqfd}) are very similar
to the well known Eilenberger \cite{Eilenberger} equations and
posses the same invariant $\ \tilde g_{1k}^2+f_{1k}^{\dagger
}f_{1k}={\rm const}.$ Now we turn to the function $f_2(\tau ):$
Using the definition of the Green function (\ref {green_func_eq})
one can take the variational derivative in Eq.(\ref{f2_var})
obtaining
\begin{equation}
f_{2k}(\tau )=-Z_k^{-1} [f_k^{II}(\tau )g_k^I(\tau )+\bar g_k^{II}(\tau )f_k^I(\tau )]
\end{equation}
where ${g^I_k,f^I_k}$ are the components of the matrix Green function
\begin{equation}
\hat g^I_k(\tau )\equiv \left[
\begin{array}{cc}
g^I_k(\tau ) & f^I_k(\tau ) \\
f^{\dagger I}_k(\tau ) & \bar g^I_k(\tau )
\end{array}
\right] =\int d\tau _2 \hat G_{1k}(\tau ,\tau _2)
\end{equation}
and $f^{II}_k,\bar g^{II}_k$are the components of the matrix Green function $%
\hat g^{II}_k(\tau )=\int d\tau _1 \hat G_{1k}(\tau _1,\tau ).$ Equations that determine
the functions $\hat g^I$ and $\hat g^{II}$ can be easily found by
integrating Eq.(\ref{green_func_eq}) over $\tau _2$ and transposed
Eq.(\ref{green_func_eq}) over $\tau _1$
\begin{eqnarray}\nonumber
\label{g_I}\partial _\tau g_k^I(\tau )+{\cal H}_k(\tau )g^I_k(\tau )=1   \\
\label{g_II}-\partial _\tau g^{II}_k(\tau )+g^{II}_k(\tau ){\cal H}_k(\tau )=1.
\end{eqnarray}

{\it Initial and final conditions.}
Initial and final conditions should be formulated for the
Green function  $\hat g_k^{\alpha ,\beta }(\tau )$    which is
defined with respect to the evolution operator $S^{\dagger }S,$
$$
\hat g_k^{\alpha ,\beta }(\tau )={\frac{{<i_G|S^{\dagger }(\infty
,0)\,S(0,\tau )\hat \psi _k^\alpha \hat \psi _k^{\beta \dagger }S(\tau
,-\infty )\,|i_G>}}{{<i_G|\,S^{\dagger }(\infty ,0)\,S(0,-\infty )|i_G>}}}
$$
where a negative $\tau $ was chosen for concreteness. To find the function $%
\hat g_k(\tau )$ in the functional representation we add the source term $%
\int d\tau \mu _k^{\alpha \beta }(\tau ) \tilde \psi _\alpha (\tau )
\tilde\psi_\beta^\dagger(\tau )$ where $\tilde\psi$ is the Nambu spinor
to the Lagrangian (\ref{lagrp}) and take the variation derivative
of (\ref{final_prob}) with respect to $\mu _k^{\alpha \beta }(\tau )$
obtaining
$$
\hat g_k(\tau )=\hat g_{1k}(\tau )+\hat g_{2k}(\tau ),
$$
where $\hat g_{1k}$ is defined by Eq.(\ref{definition_g1}) and the function
$$
\hat g_{2k}(\tau )=\left[
\begin{array}{cc}
g_{2k}(\tau ) & f_{2k}(\tau ) \\
f_{2k}^{\dagger }(\tau ) & \bar g_{2k}(\tau )
\end{array}
\right]
$$
is related with $g^I$and $g^{II}$ by
\begin{equation}
 \hat{g}_{2k}=-\hat{g}^I_k \, \hat{g}^{II}_k/Z_k.   \label{g_2_g_I_g_II}
\end{equation}
 In the superconducting phase ($|\tau |\gg |T|$) the function $\hat g_k$
should coincide with the equilibrium superconducting Green
functions
\begin{equation}
\label{bc1}\tilde g_k= \xi _k/2\,E_k
,\;\; f_k=\Delta / 2 E_k
,\;\;f_k^{\dagger }=f_k^{*},\;\;s_z=0
\end{equation}
where $E_k=\sqrt{\xi_k^2+\Delta^2},$
$\tilde g_k=[g_k-\bar g_k]/2, \,s_{zk}=-[g_k+\bar g_k]/2.$ The function $
\tilde g$ is directly related with electron density on the level $k$ by $
n_k=1-2\tilde g_k$ and the function $s_{zk}$ is the z-component of the spin
on the level $k.$ Analogously, in the paramagnetic phase the Green function $%
\hat g$ is
\begin{eqnarray}
\tilde g_k&=&f_k=f_k^{\dagger }=0,\;\;\;\;\; s_{zk}=1/2\;\;\;\; { \rm for}
\;\;\;\;\;\;|\xi _k|<\tilde \xi     \label{param_states}              \\
\tilde g_k&=&{\rm sign} \xi_k/2, \;\;\; s_{zk}=f_k=f_k^{\dagger }=0\;\; {\rm for
}\;\;  |\xi _k|>\tilde \xi .   \label{sup_states}
\end{eqnarray}
In the absence of tunneling the physical Green function $\hat g$ would coincide with
$\hat g_1$ that
obeys equations (\ref{eqfd},\ref{eqgz}) conserving the quasiparticle spin.
Therefore the function $\hat g_1$ in the paramagnetic
state obeys the boundary condition (\ref{sup_states}) for any $\xi$ while in the superconducting
state it obeys the boundary conditions (\ref{bc1}), so that  $\hat g_2\to 0$
in the superconducting phase.

\begin{figure}[ht]
\includegraphics[width=3.3in]{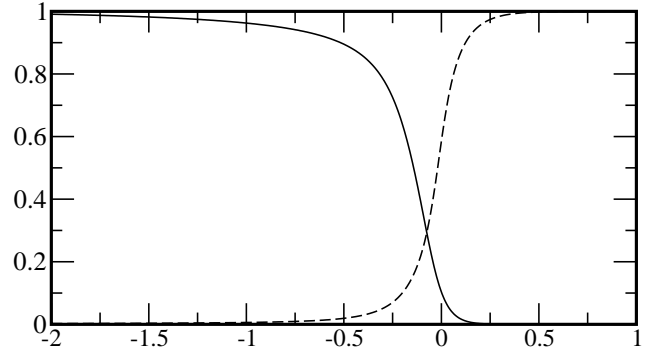}
\vspace{0.3cm}
\caption{The normalized order parameter $\Delta(\tau) / \Delta_0$ (dashed line)
and the normalized total spin $2S(\tau)/N$ of the grain
as functions of
on $\tau^\prime=\tau-T$ at the boundary $\tau\sim T$ between the superconducting
(left) and paramagnetic (right) states. }
 \end{figure}

{\it Numerical solution.} Solution of Eqs.(\ref{eqfd})
satisfying the necessary boundary conditions for a given
configuration of the order parameter $\Delta(\tau)$ can be easily
found numerically. To find the components of the function $\hat
g_2$ one first needs to solve Eqs.(\ref{g_II})
numerically and then find the components $\hat g_2$ according to
Eq.(\ref{g_2_g_I_g_II}). Knowing the functions $\hat g_1$ and
$\hat g_2$ for a given configuration $\Delta(\tau)$ one can find
the self-consistent configuration of $\Delta(\tau)$  satisfying
Eq.(\ref{self_cons_delta}). This self-consistent solution
$\Delta(\tau)$ along with the total spin of the grain $S(\tau)$ at
the right instanton boundary is presented on Fig.2 (solution at
the left boundary can be obtained as the mirror reflection of that
on the right one).
 Substituting the functions $\Delta(\tau)$
and $g^I, g^{II}$ into Eq.(\ref{probabillity}) and calculating the term ${\rm Tr} \ln[
\partial_\tau+ {\cal H}_k]$ by the method
described in \cite{LI} one obtains the result (\ref{answer}).

{\it Conclusions.}   In conclusion, we have found the probability
of the quantum transition between superconducting and paramagnetic
states of a nanometer size superconducting grain weakly coupled to
a normal metallic plate. Our result (\ref{answer}) obtained
formally in the limit of $T\to 0$ is general and remain valid at
finite temperatures as long as $T\ll \delta E$.  At higher
temperatures, $ \delta E\ll T \ll \Delta$, the characteristic
energy $\delta E$ in the exponent of Eq.(\ref{answer}) has to be
substituted by temperature $T$.  However, finding numerical
coefficient $\beta$ in this case requires a more advanced study.

We would like to thank Y. Bazaliy, Ya. M. Blanter, E.M.
Chudnovsky, D. Feldman, Y.M. Galperin, A.E. Koshelev, A.I. Larkin,
K. Matveev, A. Mel'nikov, V.L. Pokrovsky, and R. Ramazashvily for
useful discussions.
This work
was supported by the U.S. Department of Energy, Office of Science
under contract No. W-31-109-ENG-38.


\vspace{-0.6cm}

\begin{references}

\vspace{-1.6cm}

\bibitem{Grains_Experiment}  C.T. Black, D.C. Ralph, and M. Tinkham {\bf 76}
(1996) 688

\bibitem{Films_theory}  A. M. Clogston, Phys. Rev. Lett., {\bf 5} 464 (1960)

\bibitem{Films_1}  R. Meservey, P.M. Tedrow, and P. Fulde, Phys. Rev. Lett.
{\bf 25} (1970) 1270

\bibitem{Films_2}  R. Meservey, P.M. Tedrow, Phys. Rep. {\bf 238} (1994) 173

\bibitem{larkin} A.I. Larkin,
Zhurnal  Eksperimental'noi  i  Teoreticheskoi Fiziki {\bf 48} 232 (1965);
(Sov.Phys. JETP {\bf 21}, 153 (1965)).

\bibitem{Anderson}  P. W. Anderson, J. Phys. Chem. Solids {\bf 11} (1959) 28


\bibitem{Eilenberger} G. Eilenberger, Z. Phys. {\bf 214}, 195 (1968)

\bibitem{LI}  A.V. Lopatin and L.B. Ioffe, Phys. Rev. {\bf B} 6412 (1999)
\end{references}
\end{document}